\documentstyle{vciart}

\begin{document}

\input{epsf.tex}

\begin{frontmatter}

\title{EVIDENCE FOR A NARROW STRUCTURE AT W$\sim$1.68 GEV 
IN $\eta$ PHOTOPRODUCTION OFF THE NEUTRON }

\vspace{0.5  cm}
 
\author[knu,inr,cat] {V.Kuznetsov\thanksref{slava}},
\author[inr] {S.Churikova},
\author[tor] {G. Gervino}, 
\author[san] {F. Ghio},
\author[san] {B. Girolami}, 
\author[inr] {D.Ivanov},
\author[knu] {J.Jang},
%\author[inr] {D.Karapetiantz},
\author[knu] {A.Kim},
\author[knu] {W.Kim},
\author[knu] {A.Ni},
\author[inr]  {Yu.Vorobiev},
\author[knu] {M.Yurov},
\author[inr] {and A.Zabrodin}.

\address[knu] {Kyungpook Natioanal University, 1370 Sankyuk-dong, Puk-ku, Daegu, Republic of Korea}
\address[inr] {Institute for Nuclear Research, 117312 Moscow, Russia}
\address[cat] {INFN Laboratori Nazionali del Sud and Universit\`a di Catania,
95123 Catania,Italy}
\address[tor] {Dipartimento do Fisica Sperimentale, Universit\`a di Torino and INFN Sezione di Torino, 10125 Torino,Italy}
\address[san] {INFN sezione Sanit\`a and Istituto Superiore di Sanit\`a,00161 Roma, Italy}

\thanks[slava]{E-mail: Slava@cpc.inr.ac.ru, SlavaK@jlab.org}

\date{\today}

%-------------abstract----------------

\begin{abstract}

  New results on quasi-free $\eta$ photoproduction 
  on the neutron and proton bound in a deuteron target are presented.
  The $\gamma n \rightarrow \eta n$ quasi-free cross section 
  reveals a bump-like structure which is not seen in the 
  cross section on the proton. This structure may signal the existence  
  of a relatively narrow ($M\sim 1.68$ GeV, $\Gamma \leq 30$ MeV) baryon  state. 

\end{abstract}

%----------end of abstract-------------

\end{frontmatter}

Despite the availability of modern precise
experimental data, the complete spectrum of baryons is not yet well established. 
Among 43 nucleon and Delta resonances predicted 
by QCD-inspired models, almost half have yet to be
experimentally identified (``missing" resonances)\cite{pdg}.
Quantum chromodynamics may also allow for more complicated quark systems 
containing, for example, an additional quark-antiquark pair $q \bar q$ 
(pentaquarks). The existence (or non-existence) of 
this type of particles is another challenge for both theory 
and experiment.
  
Much of our knowledge on the baryon spectrum was obtained through
pion-nucleon scattering and meson photoproduction 
off the proton. Meson photoproduction off the neutron may offer a unique
tool to study certain baryons which have still not been firmly established.
Some resonances are predicted to be exclusevely photoexcited from neutrons
and not from protons\cite{hey}. For example, a single-quark transition model\cite{mok}
suggests only weak photoexcitation of the $D_{15}(1675)$ resonance 
from the proton target. On the other hand, photocouplings to the neutron 
calculated in the framework of this approach are not small. 

The possible photoexcitation of a non-strange
pentaquark state (if it exists) is of high interest as well. 
This particle is associated with the second nucleon-like member of 
an antidecuplet of exotic baryons\cite{dia,jafw}.
Evidence for the lightest member of the antidecuplet, the $\Theta^+(1540)$ baryon, 
is now being widely debated\cite{bur}. A benchmark signature of the non-strange
pentaquark could be its photoproduction on the nucleon. 
%Exact $SU(3)_F$ would forbid the proton photoexcitation into the 
%proton-like antidecuplet member. 
The chiral soliton model predicts that photoexcitation of the non-strange pentaquark 
has to be suppressed on the proton  and should occur mainly on the neutron\cite{max}. 
%even after accounting for $SU(3)_F$ violation\cite{max}. 
The mass of the non-strange pentaquark is expected 
to be near 1.7 GeV\cite{jafw,dia1,str}, with a total width of about 10 MeV 
and a partial width for the $\pi N$ decay mode, less than 0.5 MeV\cite{str}.

Among various reactions, $\eta$ photoproduction off the neutron 
is particularly attractive because
i) it selects only isospin $I=\frac{1}{2}$ final states;
ii) there is enough accurate data for the ``mirror" $\gamma p \to \eta p$ reaction;  
iii) this reaction was considered as particularly sensitive to the signal 
of the non-strange pentaquark~\cite{jafw,max,dia1,str}.
Up to now $\eta$ photoproduction off the neutron has been explored 
mostly in the region of the $S_{11}(1535)$ resonance from threshold
up to $W \sim 1.6$ GeV\cite{inc}.  
The ratio of cross sections, $(\gamma n \rightarrow \eta 
n)$/$(\gamma p\rightarrow\eta p)$, was found  
to be nearly constant, with a value near $\sim$0.67. At higher energies, 
the GRAAL Collaboration has reported a sharp rise of this ratio\cite{nstar2002}.

\begin{figure}[ht]
\centerline{\epsfxsize=5.5in\epsfbox{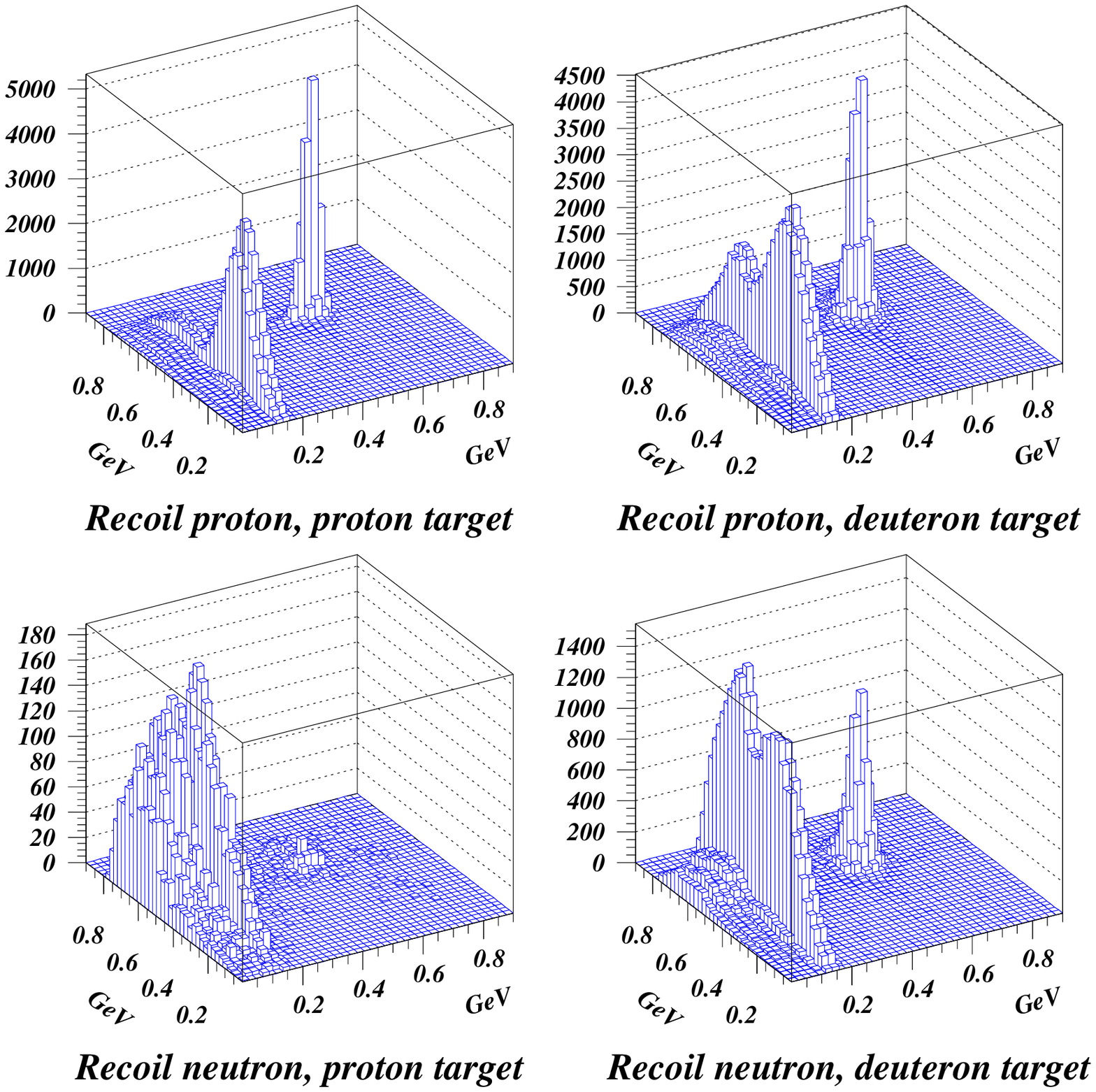}}   
\caption{ Bi-dimensional spectra of invariant mass of two photons (X axis) versus
missing mass $MM(\gamma N,N)$ calculated from momenta of recoil nucleons 
and the incoming photon (Y axis) for proton and deuteron targets. 
\label{Fig1}}
\end{figure}

In this Letter we present the analysis of data collected
at the GRAAL facility\cite{pi0} in 2002.
Both quasi-free $\gamma n \rightarrow \eta n$ and 
$\gamma p \rightarrow \eta p$ reactions were explored 
simultaneously, in the same experimental run, under the same conditions 
and solid angle using a deuteron target. 
Two photons from $\eta \rightarrow 2\gamma$ were detected in the BGO
ball\cite{bgo}. The $\eta$-mesons were identified by means of their 
invariant mass, with momentum reconstructed from the measured 
photon energies and angles. Recoil nucleons (neutrons or protons) 
were detected in two sets of detectors:\\
i) Neutrons and protons emitted at forward angles 
$\theta_{lab}\leq 23^{\circ}$, passed through two planar 
multiwire chambers, a time-of-flight (TOF) hodoscope made of thin 
scintillator strips, and a lead-scintillator sandwich TOF 
wall\cite{rw}. The latter detector provides the detection of neutrons with   
an angular resolution of $2-3^{\circ}$(Full Width at a
Half of Maximum), and a TOF resolution of $600-800 ps$ (FWHM). TOF measurement 
makes it possible to discriminate neutrons from photons 
and to reconstruct neutron momenta;\\
ii) Recoil nucleons emitted at central angles $\theta_{lab}\geq 26^{\circ}$, 
were detected in the BGO ball\cite{bgo}.  This detector provides partial 
discrimination of neutrons from photons and no TOF measurement. 
The neutron energy was obtained using kinematics constraints. 

Fig.~1 shows bi-dimensional plots of the $\gamma\gamma$ invariant mass versus 
the missing mass $MM(\gamma N,N)$ calculated from the momentum of 
the recoil nucleon (proton or neutron) and the momentum of the 
incoming photon. The plots have been obtained using data collected 
in experimental runs with proton and deuteron targets. A peak 
with coordinates ($X=m_{\eta}$, $Y=m_{\eta}$)  
corresponds to $\eta N$ photoproduction. A good $\eta p$ signal was 
obtained with the proton target, while only a few $\eta n$ events  
appeared in this run. Signals of both final states are clearly seen 
with the deuteron target.

\begin{figure}[ht]
\centerline{\epsfxsize=5.5in\epsfbox{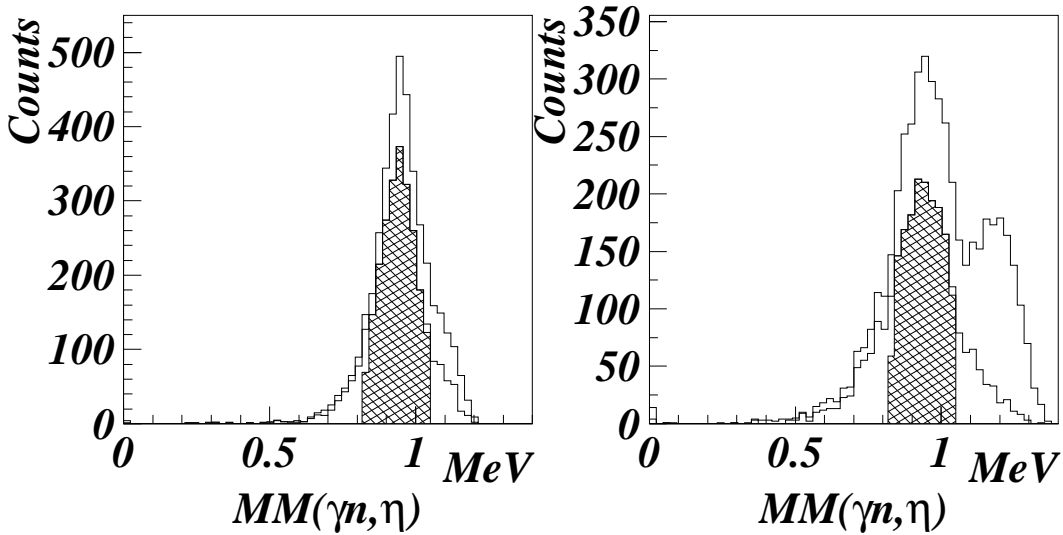}}   
\caption{Spectra of $MM(\gamma n,\eta)$ missing mass at photon energies $0.95 \leq E_{\gamma} \leq 1.2$ MeV (left panel) and $1.2 \leq E_{\gamma} \leq 1.5$ MeV (right panel). Upper curves correspond to initial selection. Lower curves indicate events after the cuts except the cut on $MM(\gamma n,\eta)$. Dashed areas show finally selected events.   
\label{Fig_mm}}
\end{figure}

As a first step of the analysis, 
the identification of the $\eta n$ and $\eta p$ 
final states was achieved in a way similar to that used in the 
previous measurements\cite{gra2} on the free proton. 
The measured parameters of the recoil nucleon were compared with 
ones expected assuming a quasi-free reaction in which the photon interacts  
with only one nucleon bound in the deuteron while the second nucleon
acts as a spectator. 

At photon energies above 950 MeV, the background
from $\gamma N \to \eta X N$ was observed. This background was clearly 
seen in the spectrum of the $MM(\gamma N, \eta)$ missing mass in which it
appeared as the second bump shifted to higher mass region from 
the position of the main peak at 0.94 GeV~(Fig.\ref{Fig_mm}). To reject this background, 
the cut on $MM(\gamma N, \eta)$ was imposed. In case of the neutron 
detection in the BGO ball, this cut was added by lower and upper limits
on the BGO signal attributed to a neutron hit
$0.014$ GeV$\leq \Delta E \leq 0.5*T_n$. The latter cut was found efficient to
discriminate between neutrons and accidental low-energy photons emitted 
as secondary particles in the detector volume, and high-energy 
photons produced in background reactions. 
 
In the case of a photon interaction with a nucleon bound in the deuteron, 
event kinematics is ``peaked" around that on a free 
nucleon. Fermi motion of the target nucleon changes the effective energy of 
photon-nucleon interaction and affects momenta of outgoing 
particles. It also complicates discrimination of the
background. Some events may suffer from re-scattering and 
final-state interaction\cite{kudr}. Such events might generate an artificial 
structure in the cross section due to specific effects like virtual 
sub-threshold meson production followed by an interaction with the 
spectator nucleon~\cite{il}. 

The goal of the second stage of the analysis was to minimize 
any influence of re-scattering, final-state interaction, or background 
contamination. Here, we used the sample of events in which 
the recoil neutrons/protons were detected in the forward detectors. 
The strategy at this stage was to study the dependence of the spectra
of selected events on cuts. The recoil nucleon missing mass $MM(\gamma N, \eta)$,
$TOF_{meas}-TOF_{exp}$, and $\theta_{meas}- \theta_{exp}$ selection windows 
were reduced by a factor 2-3.
Tight cuts preferably reject re-scattering, final-state interaction,
and the remaining background. 
They also suppress those events whose kinematics is strongly distorted by Fermi
motion or in which one or more parameters of the outgoing particles are not 
properly measured, due to detector response.

Four types of spectra were considered at this stage: \\
i)  The spectrum of the center-of-mass energy $W$ calculated from the momentum
    of the initial-state photon and assuming the target nucleon 
    to be at rest $W=\sqrt{(E_{\gamma}+M_N)^2-E_{\gamma}^2}$. 
    This quantity ignores Fermi motion and is
    peaked around the effective center-of-mass energy (40-60 MeV(FWHM) depending 
    on the energy of the incoming photon).  \\
ii)  The spectrum of the center-of-mass energy reconstructed as the
    invariant mass of the final-state $\eta$ and the nucleon
    $M(\eta N$). This quantity is much less smeared by Fermi motion (about 2 MeV(FWHM))
    but includes large uncertainties due to instrumental resolution ($40-60$ MeV(FWHM)).\\
iii) Distribution of the momentum for the spectator nucleon,
    reconstructed as the ``missing" momentum from the momenta of the final-state 
    $\eta$ and nucleon and the momentum of the incoming photon;\\
iv) Difference between the final-state $M(\eta N)$ invariant mass and
    the initial-state center-of-mass energy $W$.
 
%%%%%%%%%%%%%%%%%FIG.2%%%%%%%%%%%%%%%%%%%%%%
\begin{figure}[ht]
\vspace*{-0.9cm}
%\centerline{\epsfxsize=3.5in\epsfbox{fn2p22.jpg}}  
%\hspace{0.2 cm} \psfig{file=fn2p22.eps,width=2.2in,angle=0}
%\hspace{-0.2 cm} \psfig{file=feta2p22.eps,width=2.2in,angle=0}
\centerline{\epsfverbosetrue\epsfxsize=13cm\epsfysize=14cm\epsfbox{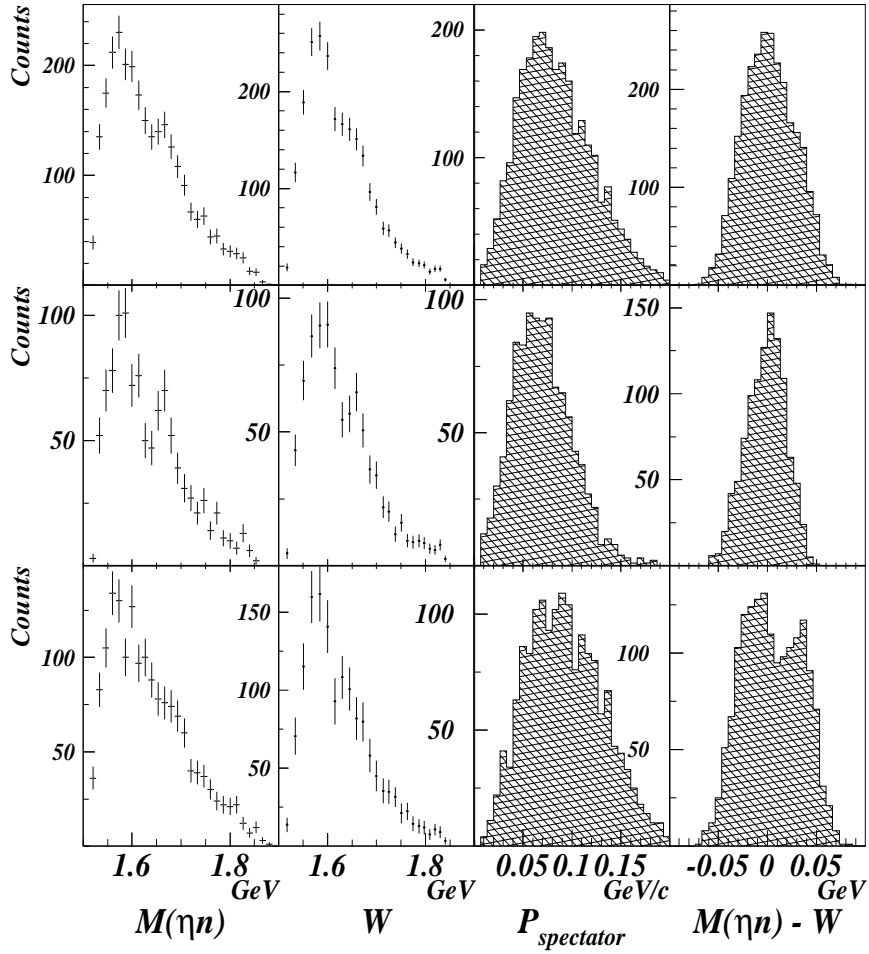}}
%\hspace{0.1cm}{\epsfverbosetrue\epsfxsize=6.85cm\epsfysize=7.1cm\epsfbox{feta2pict22.eps}}

\vspace*{0.2cm}
\caption{$\gamma n \rightarrow \eta n$ data.
         Spectra of center-of-mass energy, calculated as 
         invariant mass of final-state $\eta$ and the nucleon (left 
         columns), from the energy of the incoming photon and assuming 
         the target nucleon to be at rest (second columns), momentum 
         of the spectator nucleon (third columns), and  difference between 
         final-state and initial-state center-of-mass energies(fourth 
         columns). Upper rows correspond to initial selection, middle 
         rows show spectra after tight cuts, lower rows 
         show events rejected by tight cuts.  
         \label{fig:g1}}
\end{figure}
%%%%%%%%%%%%%%%%%%%%%%%%%%%%%%%%%%%%%%%%%%%%

The upper row of
Fig.~\ref{fig:g1} shows the  $M(\eta n)$  
(first column) and  W (second column) spectra 
obtained with the initial cuts. Both exhibit a shoulder-like 
bump in the region of 1.6 - 1.7 GeV on the slope of the $S_{11}(1535)$ 
resonance. The spectator-momentum (third column) and the
$M(\eta n) - W$ distributions (fourth column) are relatively broad. 
Plots in the middle row correspond to the tight cuts. 
Here the spectator-momentum spectrum is more compressed. The $M(\eta n) - W$ 
spectrum is more narrow and is localized near 0. The bumps observed 
in the previous $M(\eta n)$ and $W$ spectra, become more pronounced and 
are transformed into peaks near $1.68$ GeV. Conversely, events rejected 
by the second-level cuts (lower row) form a broader spectator-momentum 
distribution with the maximum near 0.1 GeV/c. The $M(\eta n) - W$ 
difference contains two maxima, both shifted from 0. The $M(\eta n)$ and $W$
spectra show some hints on lateral peaks.

%%%%%%%%%%%%%%%%%FIG.3%%%%%%%%%%%%%%%%%%%%%%
\begin{figure}[ht]
\vspace*{-0.9cm}
%\centerline{\epsfxsize=3.5in\epsfbox{fn2p22.jpg}}  
%\hspace{0.2 cm} \psfig{file=fn2p22.eps,width=2.2in,angle=0}
%\hspace{-0.2 cm} \psfig{file=feta2p22.eps,width=2.2in,angle=0}
\centerline{\epsfverbosetrue\epsfxsize=13cm\epsfysize=14cm\epsfbox{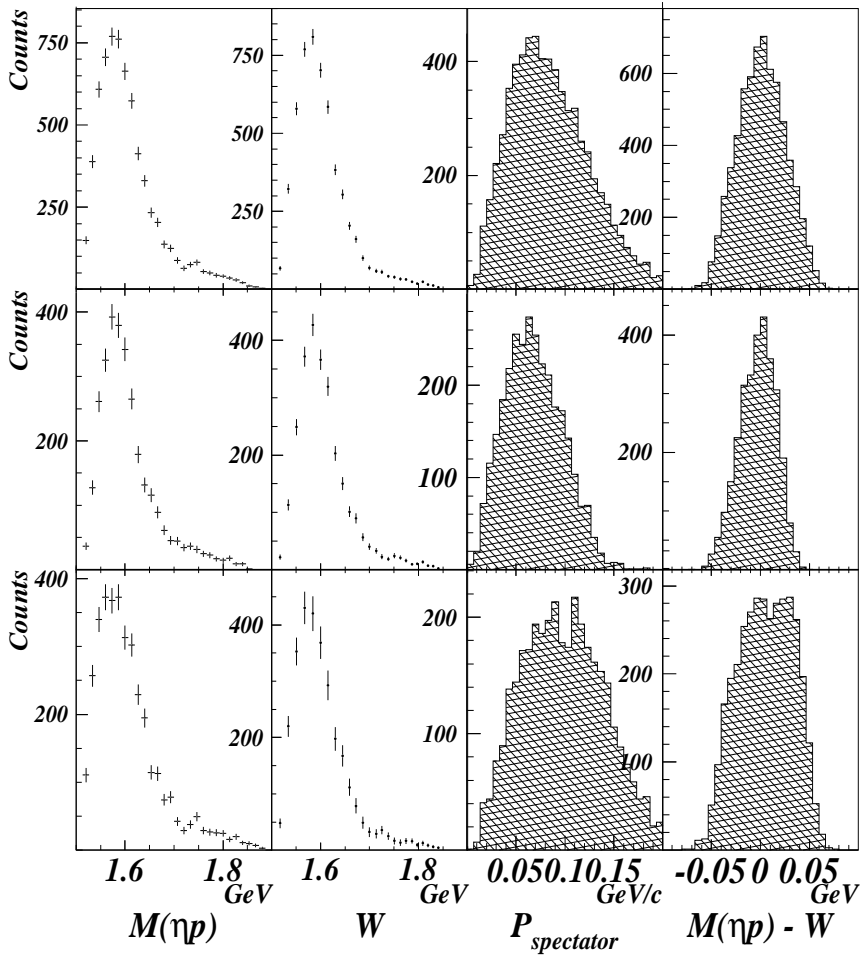}}
%\hspace{0.1cm}{\epsfverbosetrue\epsfxsize=6.85cm\epsfysize=7.1cm\epsfbox{feta2pict22.eps}}

\vspace*{0.2cm}
\caption{$\gamma p \rightarrow \eta p$ data.
	The legend is the same as for the Fig.~\ref{fig:g1}.
         \label{fig:g1a}}
\end{figure}
%%%%%%%%%%%%%%%%%%%%%%%%%%%%%%%%%%%%%%%%%%%%

The same procedure was applied to the quasi-free $\gamma p \to \eta p$ reaction 
(Fig.~\ref{fig:g1a}). The spectator momentum and $M(\eta p) - W$ 
spectra are similar to those obtained on the neutron. However, the $M(\eta N)$ and $W$ 
spectra are smooth and exhibit no structure. 

Evolution of spectra in Fig.~\ref{fig:g1},~\ref{fig:g1a} suggests that most of events rejected 
by the second-level cuts either strongly suffer from Fermi motion and/or
detector response, or possibly originate from re-scattering and 
final-state interaction. However, events shown in the middle-row 
plots, correspond to quasi-free reactions. These spectra clearly reveal 
a peak at 1.68~GeV in $\eta$ photoproduction on the neutron 
which is not seen on the proton.

%%%%%%%%%%%%%%%%%FIG.4%%%%%%%%%%%%%%%%%%%%%%
\begin{figure}[ht]
\vspace {-0.2cm}
%\centerline{\epsfxsize=5.0in\epsfbox{fn2p41b.eps}}  
%\centerline{
%\psfig{file=f0a.eps,width=4.0in,clip=0,silent=,angle=0}
%}
{\centerline\epsfverbosetrue\epsfxsize=14.0cm\epsfysize=10.cm\epsfbox{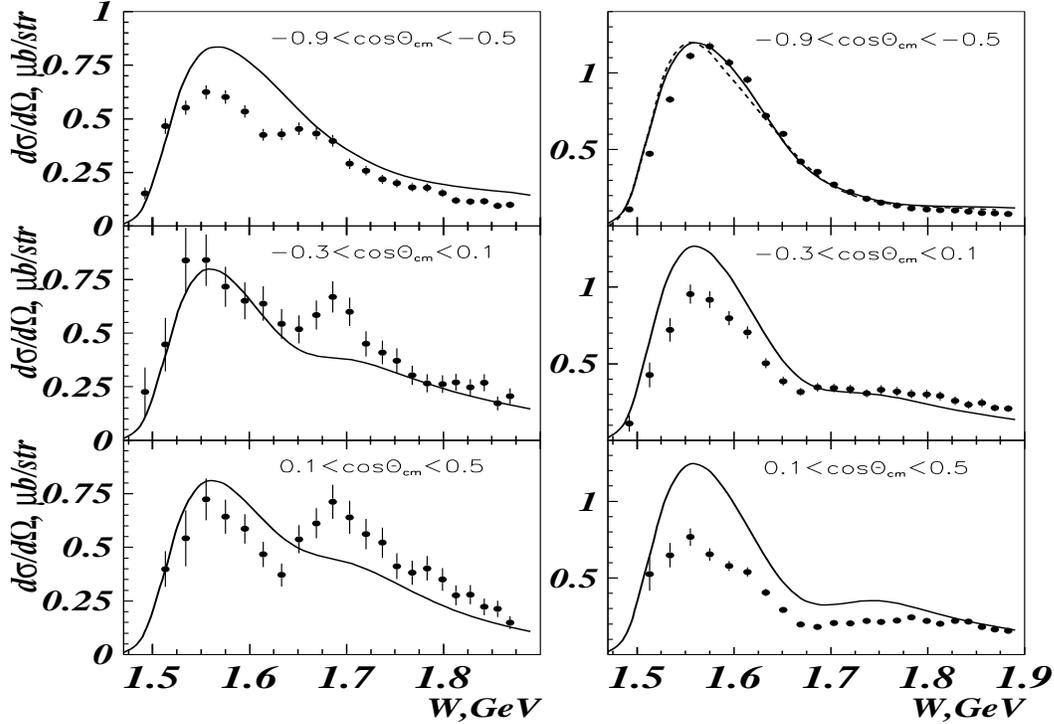}}
%\vspace*{2.3 cm}
\caption{ Quasi-free differential cross-section
         at different angles .
         Left panel:  $\gamma n \rightarrow \eta n$. 
	 Right panel: $\gamma p \rightarrow \eta p$.
	 Solid lines are $\eta$-MAID
         predictions for $\eta$ photoproduction on the free neutron/proton 
	 folded with Fermi motion. Dashed line is E429 solution of the SAID 
	 $\gamma p \to \eta p$ partial wave analysis folded with 
	 Fermi motion. 
          \label{fig:cr2}}
%\vspace*{-0.5cm}
\end{figure}
%%%%%%%%%%%%%%%%%%%%%%%%%%%%%%%%%%%%%%%%%%%%

The measured quasi-free differential cross sections for $\eta n$ and $\eta p$ 
photoproduction are shown in Fig.~\ref{fig:cr2}. 
The common normalization for both protons and neutrons was done by comparing 
quasi-free proton data at backward angles with the E429 solution of the SAID $\gamma 
p\to\eta p$ partial-wave analysis\cite{str2} and $\eta$ - MAID prediction\cite{maid}
for $\eta$  photoproduction on the free proton, which were folded with Fermi
motion (upper row, right panel of Fig.~\ref{fig:cr2}). 
The measured spectra of events were corrected on the simulated detection efficiency 
and on the beam spectrum. 
In addition the spectra of $\gamma n \to \eta n$ events were corrected on the difference 
between the measured and simulated efficiencies of the neutron detection.
The neutron detection efficiency was determined using the previous data 
for the $\gamma p \to \pi^+ n$ reaction\cite{gra1}. 
It was found to be about 22\% for the shower wall 
and 27\% for the BGO ball being dependent the neutron energy, on the pulse height 
thresholds set for both detectors, and on cuts used to identify neutrons. 
The obtained distributions were then scaled by a common constant factor.
The latter was determined requesting the minimum of the difference
between quasi-free proton data at backward angles and the SAID and MAID solutions.
The region of backward angles was chosen for the normalization 
because of the coincidence in shapes of 
the cross section on the proton and the SAID and MAID solutions 
(top right panel of Fig.~\ref{fig:cr2}). This coincidence hints a small role
of nuclear effects at these angles. At more forward angles, 
re-scattering and final-state interaction seem to become more significant 
reaching $\sim30$\% in the region of the $S_{11}(1535)$ resonance.     
Error bars shown in Fig.~\ref{fig:cr2} correspond to  
statistical uncertainties only. The normalization uncertainty 
of 10\% originates mostly from the quality of simulations of quasi-free processes 
and from uncertainties in determining the neutron detection efficiency.

%%%%%%%%%%%%%%%%%FIG.5%%%%%%%%%%%%%%%%%%%%%%
\begin{figure}[ht]
%\vspace*{0.7  cm}
\centerline{\epsfverbosetrue\epsfxsize=13.5cm\epsfysize=11.5cm\epsfbox{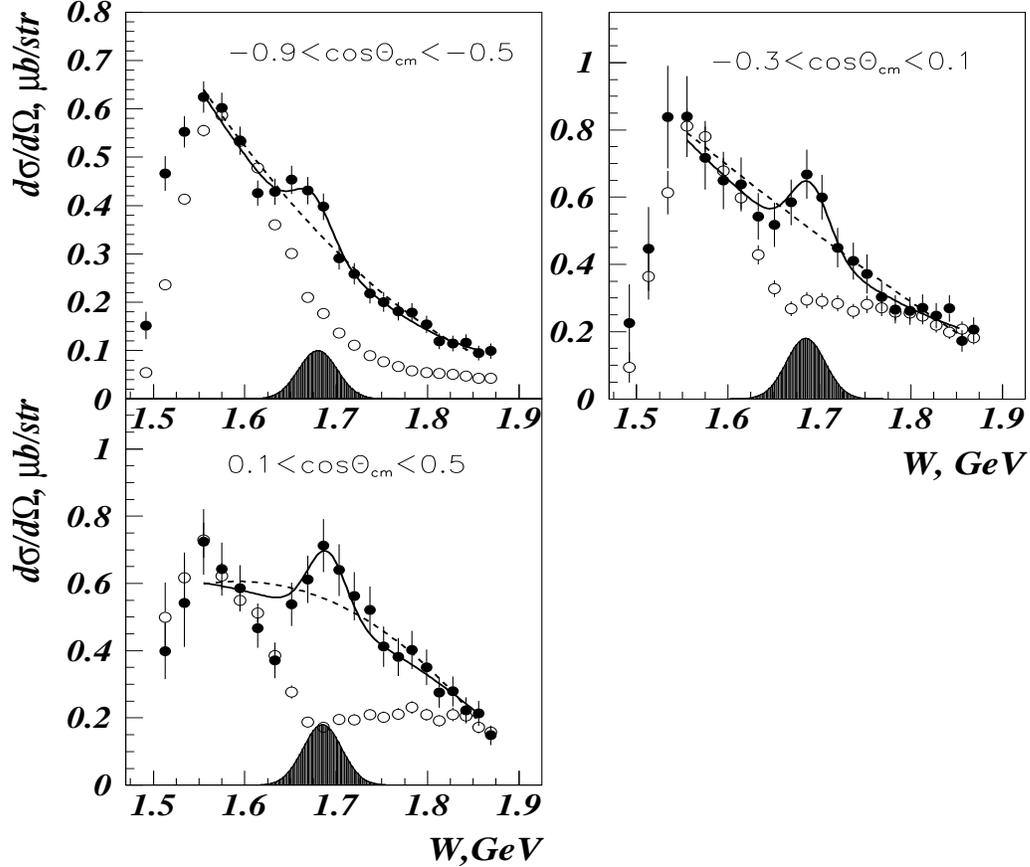}}

\caption{Polynomial-plus-narrow-state fit of $\gamma n \to \eta n$ cross sections.
Black circles are $\gamma n \to \eta n$ data. Open circles correspond to $\gamma p \to \eta p$
cross section normalized on the cross section on the neutron in the maximum of the 
$S_{11}(1535)$ resonance. Dashed areas show simulated contribution of the narrow state. 
Solid lines are the result of the fit. Dashed lines show the fit by 3-order polynomial only. 
\label{fig:cr3}}
%\vspace{-0.3cm}
\end{figure}
%%%%%%%%%%%%%%%%%%%%%%%%%%%%%%%%%%%%%%%%%%%%%%%%%%%%%%%%%%%%%%%%%%%%%%

The cross section on the neutron clearly reveals 
a bump-like structure\footnote{ The cross section 
obtained with tight cuts exhibit a slightly more narrow structure 
but includes larger statistical and systematic errors. 
For the sake of clarity and reliability in our conclusions it is not shown.}
near $W \sim 1.68 $ GeV. This structure looks slightly wider at forward angles. 
The visible width of the peak at forward angles is about $80-100$ MeV(FWHM)
(or $rms= 35 - 40$ MeV). 
The data have been compared with an isobar model for 
$\eta$ photo- and electroproduction $\eta - MAID$\cite{maid}. 
The model includes 8 main resonances and suggests the dominance of the
$S_{11}(1535)$ and $D_{15}(1675)$ resonances in $\eta$ photoproduction
off the neutron below $W\sim 1.75 $ GeV.   
The model predicts a bump-like structure near $W \sim 1.7$ GeV
in the total $\eta$ photoproduction cross section on the
neutron\cite{tia}. This structure is caused by the $D_{15}(1675)$ resonance. 
The $\eta$ - MAID differential cross sections  are  
smooth (Fig.~\ref{fig:cr2}, left panel). The PDG estimate for the $D_{15}(1675)$ $\eta N$ 
branching ratio $\frac{\Gamma_{\eta N}}{\Gamma_{total}}$ is close to 0 while the value included into
$\eta$-MAID is 17\%\cite{tia}. The PDG average 
for the Breit-Wigner width of this resonance is 
$\Gamma \sim 150$ MeV\cite{pdg}. The structure observed in the quasi-free cross section
looks  more narrow. 
%Comparison of $\eta$-MAID with beam asymmetry data 
%is ambiguous (Fig.~\ref{fig:sig}).

It is well known that $\eta$ photoproduction on the
proton is dominated by photoexcitation of the $S_{11}(1535)$ resonance up to $W \sim 1.68 $ GeV.
At higher energies, the increasing role of higher-lying resonances is expected \cite{gra2,etap}.
$\eta$ photoproduction on the neutron is dominated by the $S_{11}(1535)$ 
up to $W \sim 1.62$ GeV\cite{inc,nstar2002}. The shape of cross sections on the neutron and
on the proton in the region $S_{11}(1535)$ resonance below $W\sim 1.62$ GeV is similar 
(Fig.~\ref{fig:cr3}). One may assume that the enhancement in the cross section on the neutron
at $W\sim 1.62 - 1.72 $ GeV is caused by an additional relatively narrow resonance. 
In Fig.~\ref{fig:cr3} the simulated contribution of a narrow state ($M\sim1.68$ GeV, $\Gamma = 10$ MeV) 
is shown. This state appears as a wider bump in the quasi-free cross section due to
Fermi motion of the target neutron. 
The neutron cross section in the range of $W\sim 1.55 - 1.85$ GeV is well fit by the sum 
of a third-order polynomial and a narrow state, with an overall $\chi^2$ about 11/14, 8/14 and 11/14 for the backward, central and forward angles respectively. The fit by only a third-order polynomial increases $\chi^2$ to about 31/15, 21/15, and 23/15.

Thus, the apparent width of the structure in the $\gamma n \rightarrow \eta n$ cross
section is not far from one expected due to smearing by Fermi motion.
The same structure was observed in the $M(\eta n)$ invariant mass spectra(Fig.~\ref{fig:g1}). 
The width of the peaks in the $M(\eta n)$ spectra is also close to experimental 
resolution. Therefore this structure may signal the existence of a relatively 
narrow $(\Gamma \leq 30$ MeV) state. If so, its properties, the possibly narrow width and the strong photocoupling to the neutron, are certainly unusual. 
There are six well-known nucleon resonances
in this mass region\cite{pdg}: $S_{11}(1650)$, $D_{15}(1675)$, $F_{15}(1680)$, 
$D_{13}(1700)$, $P_{11}(1710)$, and $P_{13}(1720)$. Among them $D_{15}(1675)$ was predicted to have stronger photocouplings to the neutron\cite{hey,mok}. One cannot exclude that the observed 
structure might be a manifestation of one of them or might originate from the intereference 
between several resonances. On the other hand, such a state coincides with the expectation of the chiral soliton model\cite{max,dia1} and a modified PWA\cite{str} for the non-strange pentaquark\footnote{Here we note that the recent negative reports on the search for the $\Theta(1540)$ pentaquark\cite{clas} put doubts  on the existence of the exotic antidecuplet and the non-strange pentaquark.}.

The possible role of some resonances has been recently examined in Ref.~\cite{tia,kim,skl} on the base of our\cite{nstar2004} and CB-TAPS\cite{kru} preliminary reports. In the standard $\eta$-MAID model the $D_{15}(1675)$ resonance produces a bump near $W \sim 1.68$ GeV in 
the total $\eta$-photoproduction cross section on the neutron. 
The unusually large branching ratio of $D_{15}(1675)$ to $\eta N$ 
is needed to reproduce experimental data.
%This violates $SU(3)$  bounds\cite{max1} for $\beta_{\eta N}=17$ and is in contradiction
%with the PDG quotation. 
The inclusion of a narrow $P_{11}(1675)$ resonance with parameters suggested in \cite{max} into $\eta$-MAID generates a narrow peak in the cross section on the free neutron while the cross section on the free proton remains almost unaffected. The peak is transformed into a wider bump similar to experimental observation if Fermi motion is taken into account\cite{tia}. The similar result has been obtained in Ref.~\cite{kim}. Authors of \cite{skl} have demonstrated that the  peak at $W\sim1.67 GeV$ in the $\eta$-photoproduction cross section on the neutron can be explained in terms of the $S_{11}(1650)$ and $P_{11}(1710)$ resonance excitation.
% with appropriate helicity amlitudes $^{n}A_{\frac{1}{2}}$. 
%The signature of this approach is the rise of differential cross sections on the 
%neutron at backward angles in the region of $W \sim 1.64 - 1.73$ GeV.

The decisive identification of the observed structure requires a complete partial-wave analysis based on a fit to experimental data. New beam asymmetry data 
from GRAAL and cross sections from the CB/TAPS Collaboration\cite{kru} and
from Laboratory of Nuclear Sciences of Tohoku University\cite{kas} 
are expected to enlarge the data base. The problem is that such analysis requires a fit to quasi-free data smeared by Fermi motion and distorted by re-scattering and final-state interaction. 
The use of the beam asymmetry $\Sigma$ is going to be even more sophisticated: considerable theoretical effort is needed to understand the interaction of polarized photons with bound nucleons\cite{sib}. More perspective seems to search for the traces of this state in reactions 
of the free proton. Another way is to study the $\gamma n \to \eta n$ reaction
in experiments with the detection of the spectator proton, and/or 
in double-polarization experiments with parallel/antiparallel beam-target polarisations. 
A spin-1/2 state would be seen only with antiparallel (helicity-1/2) beam-target polarisations. Such dedicated experiments could be carried out at JLAB and the upgraded ELSA and MamiC facilities.

It is a pleasure to thank the staff of the European Synchrotron Radiation Facility 
(Grenoble, France) for stable beam operation during the experimental run.
We thanks Y.~Azimov, K.~Goeke, and M.~Polyakov for the valuable
theoretical contribution in support of this work, R.~Workman and I.~Strakovsky
for the assistance in preparation of a manuscript.
Discussions with W.~Briscoe, V.~Burkert, D.~Diakonov, A.~Dolgolenko, I.~Jeagle, H.-~C.Kim,
M.~Kotulla, B.~Krusche, A.~Kudryavtsev, V.~Mokeev, E.~Pasyuk, 
P.~Pobylitsa, M.~Ripani, A.~Sibirtsev, I.~Strakovsky, M.~Tauti,
L.~Tiator and R.~Workman were very helpful. This work has been 
supported by Universit\`a degli studi di Catania and Laboratori Nazionale 
del Zud, INFN Sezione di Catania (Italy), 
and by Ruhr-Universit\"at Bochum (Germany).

%%%%%%%%%%%%%%%%%%%%%%%%%%%%%%%%%%%%%%%%%%%%%%%%%%%%%%

\end{document}